# Effects of number of digits in large-scale multilateration

Jean Marc Linares[a], Santiago Arroyave-Tobon[a], José Pires[b], Jean Michel Sprauel[a]

[a] *Aix Marseille Univ, CNRS, ISM, Marseille, France*

[b] *Trescal SA, Parc d'affaires Silic, 8 rue de l'Estérel - BP 30441, 94593 Rungis Cedex, France*

**Abstract:** Since many years ago, multilateration has been used in precision engineering notably in machine tool and coordinate measuring machine calibration. This technique needs, first, the use of laser trackers or tracking interferometers, and second, the use of nonlinear optimization algorithms to determine point coordinates. Research works have shown the influence of the experimental configuration on measure precision in multilateration. However, the impact of floating-point precision in computations on large-scale multilateration precision has not been addressed. In this work, the effects of numerical errors (rounding and cancellation effects) due to floating-point precision (number of digits) were studied. In order to evaluate these effects in large-scale multilateration, a multilateration measurement system was simulated. This protocol is illustrated with a case study where large distances (≤20 m) between pairs of target points were simulated. The results show that the use of multi-precision libraries is recommended to control the propagation of uncertainties during the multilateration computation.

## 1. Introduction

In the last 40 years, large-scale measurement or dimensional metrology has been an active research field in the world. Several keynote papers have been published highlighting the scientific advances in this research field [1-3]. These works have been supported by the increase of large mechanical systems (aircraft wings, wind turbine, rotor blades, mechanical structures such as nuclear reactors…). In order to produce and control the constitutive parts of these mechanical systems, large machine tools and coordinate measuring machines are required. In turn, these mechanisms also need to be calibrated and compensated [4]. In consequence, a new measuring technique appeared: multilateration. In the context of digital enterprise, measurement systems using this technique are being used increasingly in precision engineering. Maropoulos et al. [5] defined this research field as a priority in the context of measurement-assisted assembly.

The multilateration technique is based on the computation of the coordinates of a given point using either four measurement devices simultaneously or a single device sequentially. This computation is feasible if the position coordinates of the measurement devices and the distances to the target point are known. The coordinates of the target point are calculated as the intersection of four spheres. Each sphere is centered at the measurement devices position and its radius is defined by the measured distance. The intersection of the first two spheres generates a circle. Two points can be derived by intersecting this circle and the third sphere. The last sphere allows to obtain the coordinates of the target point.

Two types of instruments are chiefly used in the multilateration technique. The first type is the laser tracker, which measures simultaneously a distance by interferometry or absolute distance-meter and two angles by encoders. The second type is the Tracking Interferometer (TI, such as laser tracer) which measures by interferometry the distance between its standard sphere and the Spherically Mounted Retroreflector (SMR) without taking into account the laser dead zone.

The multilateration technique has been used in numerous research areas such as electrical and electronic engineering, telecommunications and aerospace engineering… Around 200 research papers related to this topic can be found in the Web of Science database. In the field of machine compensation and calibration, the multilateration principle was introduced by Schwenke et al. [6] in order to calibrate a machine structure using a TI. Muralikrishnan et al. [7] wrote a survey of the literature about the use of TIs in large-scale dimensional metrology. That paper is focused on error modeling, measurement uncertainty, performance evaluation and standardization. Norrdine [8] proposed an algebraic approach to solve nonlinear problems in multilateration.

Norrdine derives the spatial coordinates of the unknown points as a function of the distances to other known points using a system of quadratic equations. In the case of sequential multilateration using a TI, the "other known points" (positions of the laser tracer) and the laser dead zone are in fact not known. However, it helps to explain the quadratic equations of a global multilateration problem. Gao et al. [9] summarized the multi-axis coordinate measurement methods such as triangulation and multilateration. To determine the TI position coordinates, Chen et al. [10] proposed a calibration procedure based on additional measurements comparatively to the classic multilateration procedure. Camboulives et al. [11] presented a calibration procedure of a 3D working space based on multilateration using only one TI. The reference datum system used in this procedure was built from the successive locations of a single TI independently of the machine reference frame.

A considerable amount of research has been done on the multilateration using laser trackers. The impact of variations in system configuration (laser tracker positions) on the volumetric measurement error was studied by Zhang et al. [12]. In their article, the authors propose an optimization of the system configuration and a self-calibration planning to reduce the propagation of volumetric measurement errors. This reduction was obtained by increasing the number of laser tracker stations. In the same way, a new procedure to calibrate an articulated arm coordinate measuring was presented in that paper. Wang et al. [13] used a genetic algorithm to optimize the laser tracker positions of multilateration measurements. Santolaria et al. [14] proposed a self-calibration algorithm of four laser trackers. The same research team presented a work about the different techniques and factors that affect the measurement accuracy of laser trackers used in machine tool volumetric verification [15] and proposed different calibration strategies based on network measurements [16]. Recent papers propose the calibration of coordinate measuring machines [17] and machine-tools [18-20] using this measurement principle.

In short, most of the research works about multilateration precision are centered around the influence of the experimental configuration (the positions of the measurement device and of the target points as well as the uncertainties of the temperature, pressure and humidity sensors) on measure precision [21-22]. However, the effects of the number of digits used during computations in large-scale multilateration precision have not been addressed [23].

In response to this shortcoming, the aim of this paper is to bring to the fore the impact of the number of digits used in computation on large-scale sequential multilateration using a TI. In order to do this, two numerical experiments simulating multilateration-based measurements using a TI were performed. These experiments were conducted using the multilateration model and the measurement configuration described in Section 2. The first experiment, which is detailed in Section 3, aimed to determine the impact of numerical errors in the course of the nonlinear least-squares algorithm used in multilateration. This was evaluated by solving the floating-point calculations of the optimization problem using different numbers of digits (10 to 20). The obtained results were compared with the nominal solution of the problem. The second experiment, detailed in Section 4, was aimed to evaluate the combined effects of the measuring uncertainties together with numerical errors.

## 2. Multilateration model

The multilateration problem requires the implementation of a system of nonlinear equations and an optimization algorithm (in blue color in Figure 1). The implementation used in this work is presented Section 2.1. In order to find a numerical solution for a given problem (in orange color in Figure 1), setup values for the computed quantities (in green color in Figure 1) are required. The process to define the setup values in this work is presented in Section 2.2. These values and the measured quantities are encoded by the computer using the floating-point standard (in green color in Figure 1). The numbers of digits used in this work to encode these numerical quantities are presented in Section 2.3. In Section 2.4, the studied configuration is detailed.

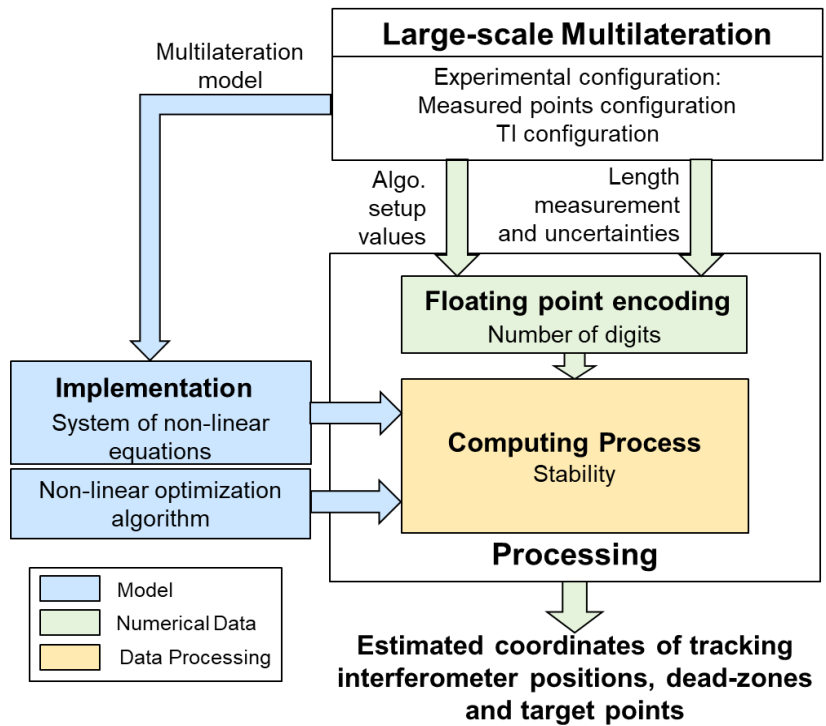

**Figure 1:** Relationship between computing process stability, number of digits and numerical results in multilateration.

*2.1 Multilateration model implementation*

In multilateration, the number of equations is linked both to the number of TI stations (*POS*) and the number of target points (*PTS*). The unknowns are the dead zones of the lasers, the coordinates of the positions of the TIs and the coordinates of the target points. The minimal number of equations is obtained when the target point coordinates are expressed in a reference frame built from the position of three TIs. In this configuration, the number of unknowns is equal to 4 *POS*+3 *PTS*-6 and the number of equations is *POS PTS*. This configuration gives a minimal number of equations (35 equations) for 7 target points and 5 TI positions. In this case the number of degrees of freedom (*dof*) is zero. In metrology or precision engineering, the number of target points is chosen to increase the number of *dof*, which implies a decrease of the uncertainty of the computed quantities. In this situation, a nonlinear optimization is performed to compute the unknowns.

In this work, this nonlinear optimization problem was implemented and solved using NonlinearFit Maple software's function. This function is based on a nonlinear least-squares algorithm.

*2.2 Setup values for the nonlinear least-squares algorithm*

In order to run the nonlinear least-squares algorithm used in multilateration, setup values (or starting points) are required. These values (nominal point coordinates, TI coordinates and laser dead zones) were randomly chosen for each simulation following the algorithm proposed in Figure 2.

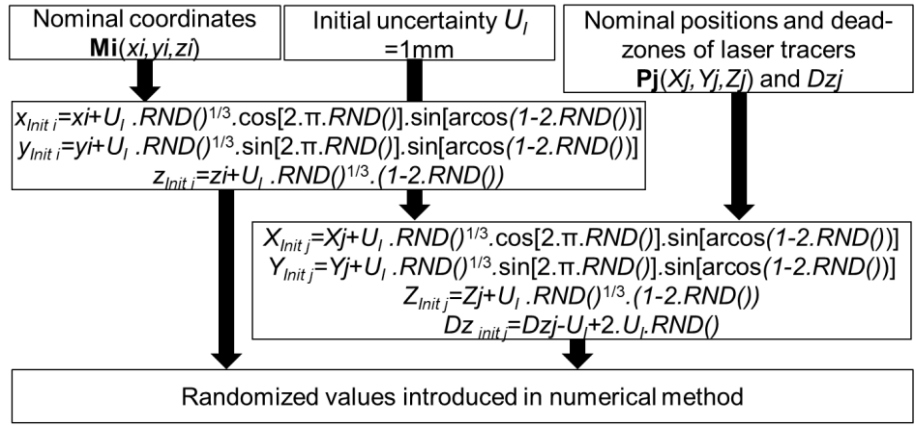

**Figure 2.** Algorithm to randomize setup values introduced in the numerical method

The randomized values were obtained from a random sampling (using the pseudo-random number generator RND() function). This sampling was uniformly distributed in an uncertanty sphere of radius $U_I$.

*2.3 Number of digits*

The nonlinear least-squares algorithm used to derive the target point coordinates in multilateration uses numerical calculations based on floating-point operations (see Figure 1). As mentioned in [23], when an algebraic expression on metrology or precision problems is well conditioned, one can always find a stable computational process to evaluate it. In contrast, it is difficult to find a stable procedure to evaluate it when it is poorly conditioned. In the case of large-scale applications for multilateration-based measures, increasing the number of target points allows to improve the stability of the computing process. However, the nonlinearity of the equations and the great difference in the order of magnitude between the measured distances (tens of m) and the reached accuracy (some μm) can generate cancellation effects during the computation.

In this work, a multilateration measurement system was simulated and the obtained equations were solved using various numbers of digits to encode real numbers. The number of digits was controlled by means of the multi-precision function 'Digits' of the Maple software. The configuration of the studied multilateration measurement system is detailed hereafter.

*2.4 Studied configuration*

To test the effect of the number of digits on multilateration computations, the measurement system depicted in Figure 3 was simulated. 11 points **Mi** (with i=1:11) were disposed in a closed space of 22m x 4m x 3m. A maximal distance of 20 m between the target points **M1** and **M11** is brought about by this configuration to reproduce measures over aircraft wings. In order to estimate the respective distances between pairs of these 11 points, five TI stations **Pj** (with j=1:5) were disposed as shown in the Figure 3. Length measurements were simulated in a sequential manner, which implies successively positioning the TI at each station **Pj** and the SMR at each target point **Mi**. The nominal coordinates of these stations and their nominal dead zones are specified in Table 1. The measurements were simulated assuming a constant temperature and an efficient quality control of humidity and pressure. This configuration gives a nonlinear system of 55 equations and 47 unknowns, which gives a *dof* = 8.

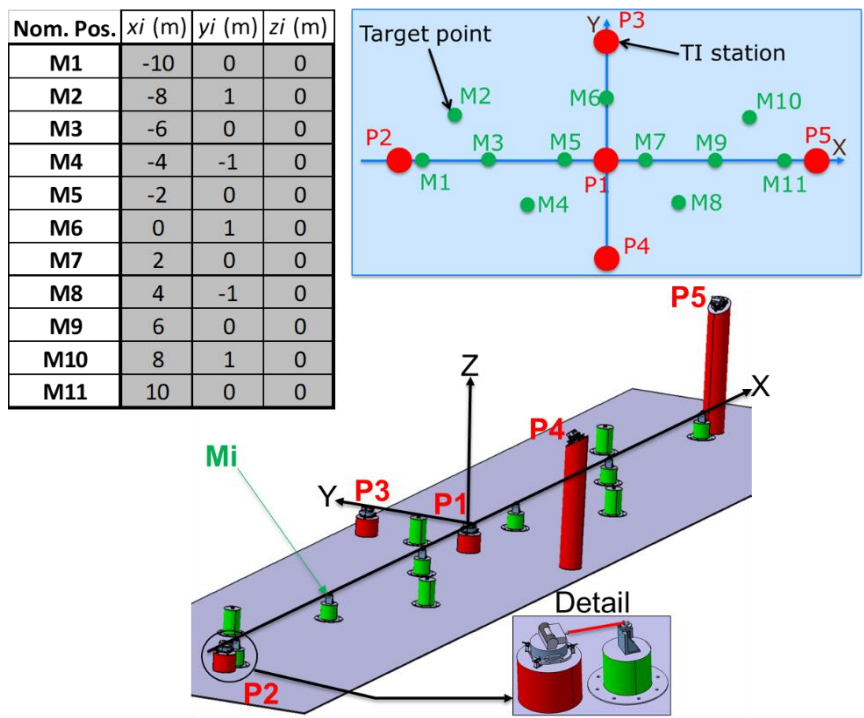

| Nom. Pos. | $x_i$ (m) | $y_i$ (m) | $z_i$ (m) |
|---|---|---|---|
| M1 | -10 | 0 | 0 |
| M2 | -8 | 1 | 0 |
| M3 | -6 | 0 | 0 |
| M4 | -4 | -1 | 0 |
| M5 | -2 | 0 | 0 |
| M6 | 0 | 1 | 0 |
| M7 | 2 | 0 | 0 |
| M8 | 4 | -1 | 0 |
| M9 | 6 | 0 | 0 |
| M10 | 8 | 1 | 0 |
| M11 | 10 | 0 | 0 |

**Figure 3.** Sets of nominal positions of the target points **Mi** (with i=1 to 11) and geometrical configuration of TI.

| Nom. Pos. | Xj (m) | Yj (m) | Zj (m) | Dzj (mm) |
|---|---|---|---|---|
| **P1** | 0 | 0 | 0 | 20.458 |
| **P2** | -10.5 | 0 | 0 | 45.455 |
| **P3** | 0 | 2.0 | 0 | 100.256 |
| **P4** | 0 | -2.0 | 2.0 | 52.230 |
| **P5** | 10.5 | 0 | 2.0 | 12.230 |

**Table 1.** Nominal positions and dead zones of TI **Pj** with j=1 to 5

## 3. Effect of the number of digits

*3.1 Simulation protocol 1*

In order to study the impact of the number of digits on sequential large-scale multilateration, the simulation protocol 1 depicted in Figure 4 was applied to the measurement configuration previously presented (Section 2.4). The impact of numerical errors during the computation was evaluated by solving the floating-point operations using different numbers of digits ranging from 10 to 20. A set of the interferometric length values was simulated using the nominal coordinate values of **Pj** and **Mi**, and without taking into account the uncertainty of the input data. The setup distance values for the nonlinear least-squares algorithm were the same for all simulations. Each simulation was performed using a different number of digits, from 10 to 20. At each computation, the computed point coordinates **Msi** were derived from the optimization algorithm. From this data, the deviations in respect to the nominal coordinates were calculated.

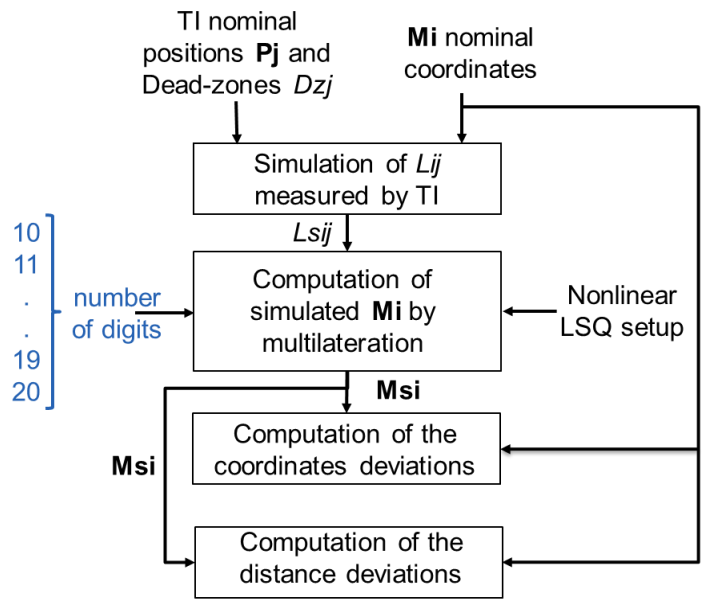

**Figure 4.** Simulation protocol to study the impact of floating-point precision in sequential large-scale multilateration where **Pj** is a nominal TI position j, *Dzj* is the dead-zone of **Pj**, **Mi** is the nominal coordinate set of target points, *Lij* is the nominal length value between **Mi** and **Pj** points, *Lsij* is the random interferometric length value, **Msi** is the computed coordinate set of target points.

*3.2 Results and discussion*

The mean value and the coverage intervals of the deviations of the computed point coordinates in respect to the nominal coordinates were calculated. The deviations reached values in the order of $10^{-6}$mm, $10^{-5}$mm and $10^{-1}$mm, for the X, Y and Z coordinates, respectively. These results may be due to the fact that the TI stations were well positioned in X and Y directions relatively to the target points. However, in Z direction the TI stations were localized only at one side of the target points, which was a physical constraint of the measurement configuration. Due to the large-scale distances measured along the X direction, the impact of the deviation of the Z coordinate over the measured distances is negligible (cosine error). The deviations of the measured distances reached values in the order of $10^{-6}$mm.

Figure 5 presents the mean of the magnitude values of the deviation vectors of all points relatively to the number of digits. A reduction of the numerical error can be noticed when the number of digits increases. The smallest computation error is obtained at 20 digits. An unexpected behavior appears for 16 and 17 digits. One can imagine different causes for this phenomenon: the encoding process, the optimization algorithm, condition number…

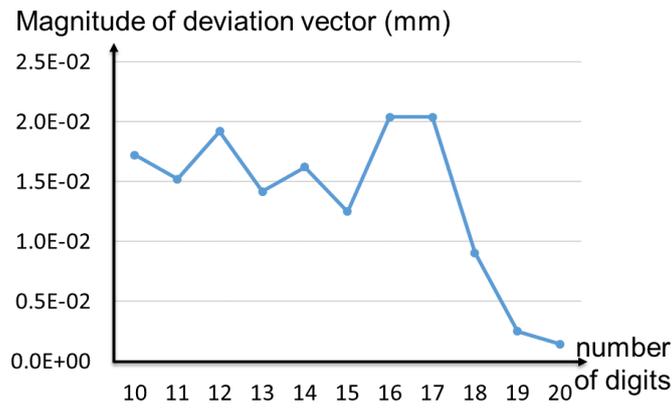

**Figure 5.** Mean of the magnitudes of the deviation vectors of all points vs number of digits.

This kind of test can be useful in precision engineering in order to determine, before performing the real computations, the most appropriate number of digits required to improve the accuracy of the computed quantities.

### 4. Effect of the number of digits considering measurement uncertainties

In order to evaluate the effects of the number of digits during multilateration computations when considering uncertainties, a second simulation protocol was defined. This protocol, summarized in Figure 6, is derived from the protocol 1, but measurement uncertainties were added. The determination of these uncertainties is detailed in Sections 4.1 and 4.2. The deviations of the simulated distances between the points were evaluated using 10 and 20 digits. The choice for these numbers of digits was motivated by the results of the previous section.

The deviations of the calculated distances were obtained as the difference between the distances (calculated from the coordinates of the simulated points **Msi**) and the nominal distances (calculated from the nominal coordinates of the target points **Mi**). These distance deviations were also computed with and without measurement uncertainties (in blue color in Figure 6). Uncertainties in distance measurements (SMR position, index of refraction (Edlen equation [24,25]) in green color in Figure 6) were set by means of a random sampling (in green color in Figure 6). All distance deviations were calculated at each loop of the simulation. 11 target points were measured in this study. The number of computed distances is 10 (M1Mi with i=2 to 11). Four experiments were defined as shown in Table 2.

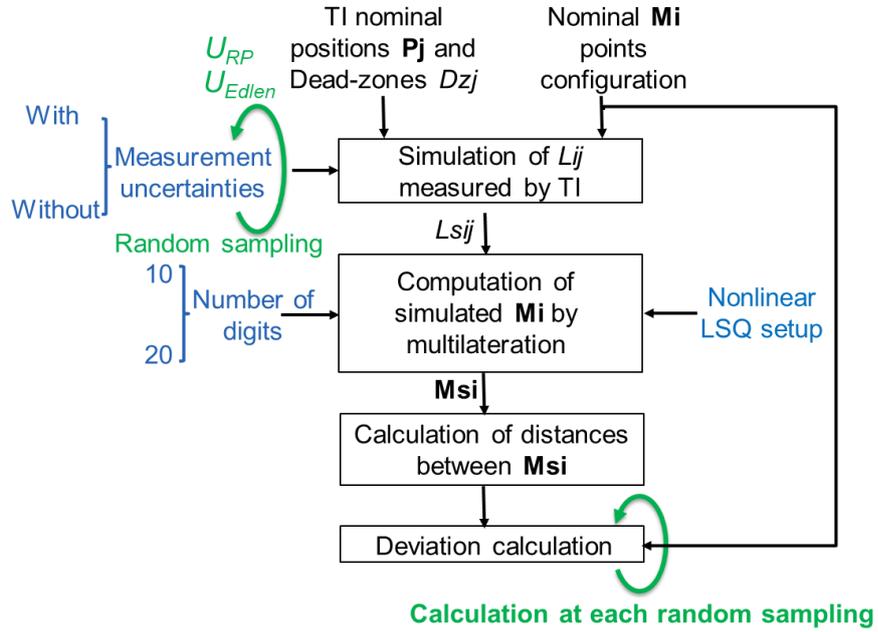

**Figure 6.** Simulation protocol to study the impact of the number of digits in sequential large-scale multilateration. **Pj** is nominal TI position j, $D_{zj}$ is the dead-zone of Pj, **Mi** is the nominal coordinate set of the *i*-th target point, $L_{ij}$ is the nominal length value between **Mi** and **Pj** points, $L_{sij}$ is the random interferometric length value, **Msi** is the coordinate set of the simulated point i, $U_{RP}$ is the SMR Position uncertainty, $U_{Edlen}$ is the interferometric length uncertainty and $U_I$ is the initial value used in nonlinear Least-squares (LSQ) algorithm.

|      | Digit nb | Uncertainties |
|------|----------|---------------|
| Exp1 | 20       | without       |
| Exp2 | 20       | with          |
| Exp3 | 10       | without       |
| Exp4 | 10       | with          |

**Table 2.** Experiments of the simulation protocol 2

*4.1. Interferometry measurement simulation*

Figure 7 depicts the algorithm employed to simulate the lengths measured by the TI for the 4 experiments. Sets of lengths including Edlen and SMR position uncertainties were simulated. A spherical random sampling was used to simulate the deviations of the SMR position $U_{RP}$ from the nominal coordinates **Mi**. A uniform repartition was imposed for this sampling. The simulated length $L_{sij}$ measured by the TI was derived from the simulated coordinates of the target points, the nominal coordinate of the TI **Pj** and the Edlen uncertainty $U_{Edlen}$. The random sampling of the Edlen uncertainty was generated following a uniform distribution. The measured lengths were simulated five times from the same TI station. The mean length of these five measures was used in the multilateration.

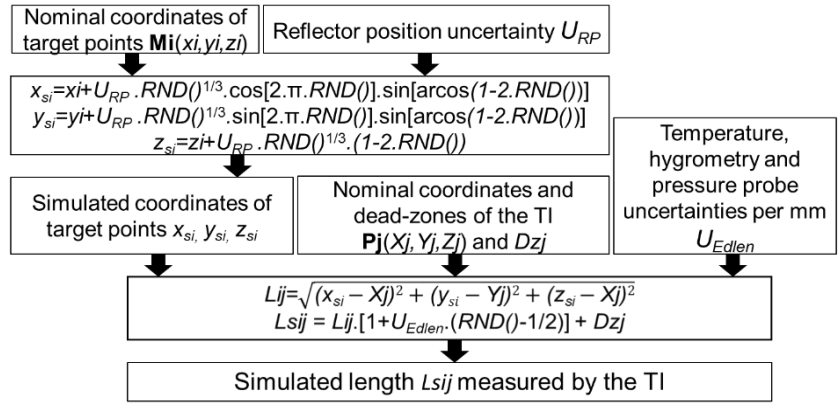

**Figure 7.** Algorithm to simulate lengths measured by TI

*4.2 Budget of measurement uncertainties*

The main contributors to uncertainty in sequential multilateration are:

- the uncertainty of the SMR position ($U_{RP}$). This uncertainty is defined by the positioning uncertainty $\Delta 1$, optical aberrations and coaxiality errors $\Delta 2$,
- the interferometric length uncertainty ($U_{Edlen}$) generated by uncertainties of the atmospheric sensors (temperature $U_t$, humidity $U_f$ and pressure $U_p$).

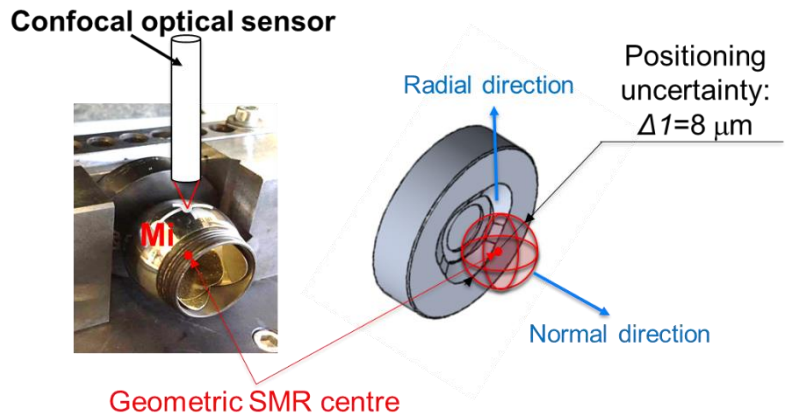

**Figure 8.** Estimation of the SMR positioning uncertainty

*4.2.1 SMR position uncertainty*

Experiments were done to determine the SMR position uncertainty $U_{RP}$ (Figures 8 and 9).

The positioning variations between the external sphere of the SMR and its support were measured ten times in the radial and normal directions. These two sets of measures were carried out using a confocal optical sensor (left-hand side of Figure 8). The maximal value of the standard deviation of both sets of measures was selected. The positioning uncertainty $\Delta 1$ was calculated with a confidence level of 95%. This value $\Delta 1$ was estimated to ±8 μm.

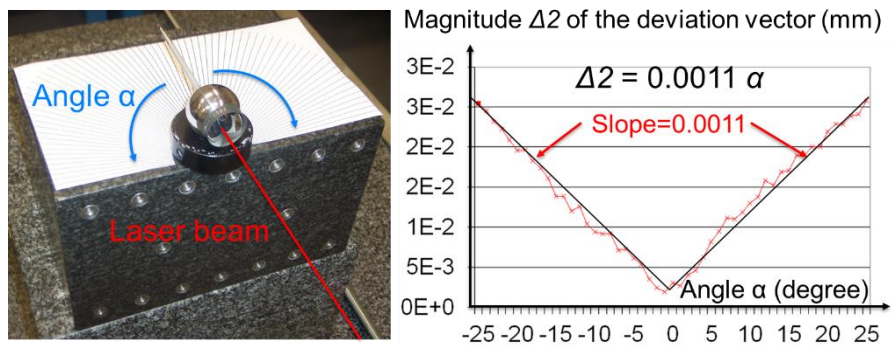

**Figure 9.** Optical aberration and coaxiality (center of the physical sphere and the optical center) uncertainty estimation

Typically, during measurements with TI, the SMR is manually aligned with the laser beam and one edge of its optical corner cube is placed vertically to minimize the optical aberration. Experiments were carried out to determine the uncertainty $\Delta 2$ arising from the alignment between the optical corner cube and the laser beam $\alpha$ in Figure 9. The position variation of the SMR was measured using a laser tracker while varying the $\alpha$ value. The obtained results are shown on the right-hand side of Figure 9.

In our simulation, a maximal misalignment of $\alpha= \pm 2.5°$ between the laser beam and the SMR was assumed. Using the relation between $\alpha$ and $\varDelta 2$ presented in Figure 9, the assumed misalignment gives a position uncertainty $\varDelta 2= \pm 2.7$ µm. The global uncertainty of the SMR position was computed as the root of the sum of the squares of both values ($\varDelta 1$ and $\varDelta 2$), and its value is $U_{RP} = \pm 8.44$ µm.

*4.2.2 Interferometric length uncertainty*

The air refractive index has a great influence on the wavelength of the interferometer laser. Metrology Institutes have studied this phenomenon and propose a modified Edlen formulae [24,25] to calculate it. This equation is used to derive the length uncertainty per meter. This length uncertainty was named $U_{Edlen}$ and estimated to $\pm 0.825$ µm/m. The procedure applied to compute this value is summarized in Figure 10. For this computation, the considered contributors were the uncertainties of the environment sensors. These uncertainties were taken from the technical characteristics of commercially available sensors. For the temperature sensor, this value was $\pm 0.2$ °C, $\pm 0.8$ %RH for the humidity sensor and $\pm 150$ Pa for the atmospheric pressure sensor.

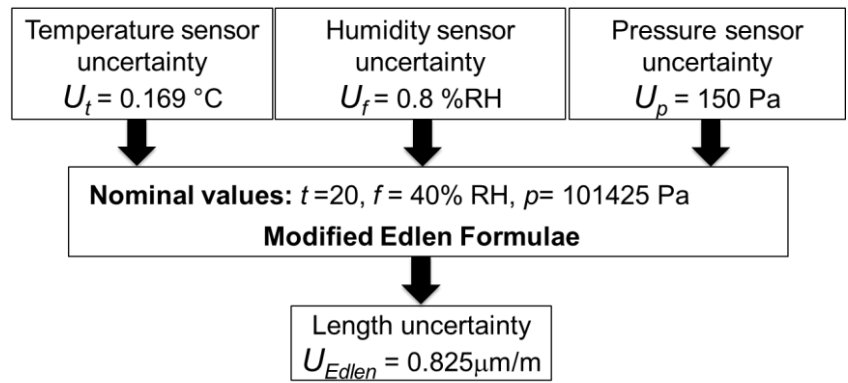

**Figure 10.** Estimation of length uncertainty ($U_{Edlen}$) versus temperature ($U_t$), humidity ($U_f$) and pressure ($U_p$) sensor uncertainties. Wavelength in a vacuum in this computation is 0.632991368 µm.

Table 3 summarizes the uncertainty contributors taken into account in the calculation of the length uncertainty.

| Uncertainty budget | ±U | Units |
|---|---|---|
| **SMR position ($U_{RP}$)** | 8.44 | μm |
| **Interferometric length ($U_{EDLEN}$)** | | |
| Temperature sensor ($U_t$) | 0.169 | °C |
| Humidity sensor ($U_f$) | 0.8 | % |
| Pressure sensor ($U_p$) | 150 | Pa |
| **Computation setup value ($U_l$)** | 1 | mm |

**Table 3** Uncertainty budget

4.3 Results of the simulation 2

The simulation protocol 2 presented in Figure 6 was executed 55 times for each experiment. The 10 distances between pairs of points were computed 55 times. The deviations between the nominal lengths and the estimated lengths were calculated. Coverage intervals for the 10 deviations were calculated for each of the 55 simulation runs. These coverage intervals indicate ± 2 standard deviations and represent a confidence level of 95%. Table 4 presents the largest value among the coverage intervals of the 10 distances for each experiment.

| | Digit nb | Uncertainties | Largest coverage interval (mm) |
|---|---|---|---|
| **Exp1** | 20 | without | 8.1 10$^{-7}$ |
| **Exp2** | 20 | with | 0.081 |
| **Exp3** | 10 | without | 5 10$^{-6}$ |
| **Exp4** | 10 | with | 0.079 |

**Table 4.** Experiments and the largest coverage intervals for the distance calculation

Figure 11 illustrates these results. In this figure, the obtained coverage intervals are represented as a function of the considered uncertainties and the number of digits. For each experiment, a circle represents the coverage intervals of the deviations between the values of nominal and estimated distances. The size of these circles is linked to the coverage interval value.

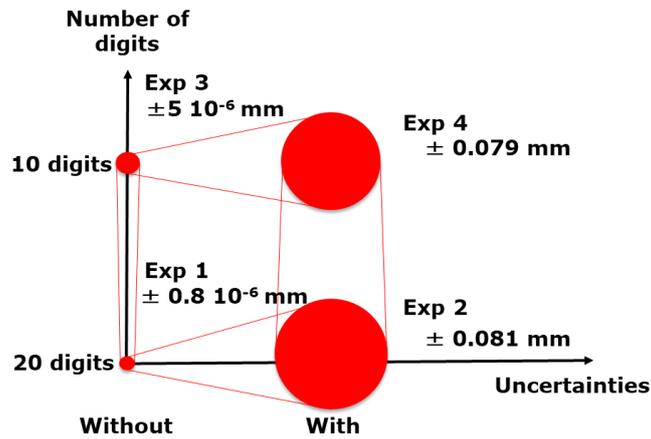

**Figure 11.** Impact of the number of digits used in computation and measurement uncertainties on distance coverage interval (mm).

From these results, it appears that the number of digits used during the computations impacts the coverage interval of the results. Without uncertainties, the results are more accurate when computed using 20 digits. The coverage interval between all distances was reduced by 84% when increasing the number of digits from 10 to 20. With uncertainties, the coverage interval obtained with 10 digits was smaller than the one calculated with 20 digits. This suggests that the rounding effects due to numerical approximations (truncation) generate an underestimation of the coverage interval. Additionally, these results suggest that the estimation of large distances using sequential multilateration can be performed with a global coverage interval value of ±0.08mm in the worst case. However, when considering the mean value of the coverage intervals of the 10 distances, the global coverage interval is ±0.046mm.

To complete the previous results, Figure 12 shows the obtained coverage intervals as a function of five distance values. The points located along the X direction were chosen to highlight the effect of distance on the obtained coverage intervals. This axis includes the largest distance (20 m) among all the points. These results are shown for both 10 and 20 digits instances (blue and red data respectively in Figure 12).

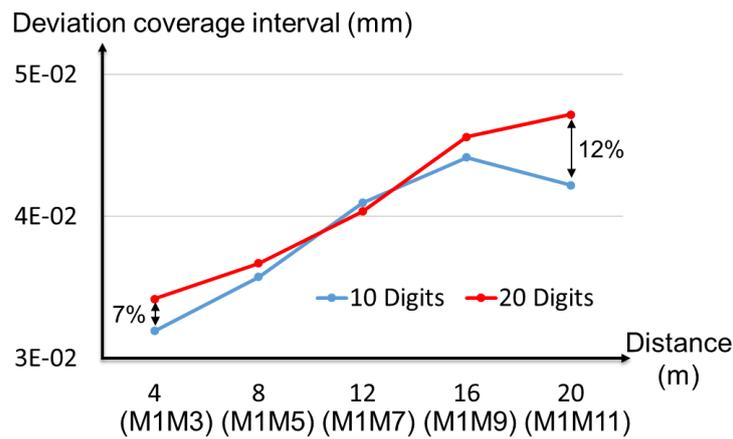

**Figure 12.** Deviation coverage interval of distances vs number of digits (10 and 20 digits).

This figure shows the same behavior, as previously discussed, regarding the underestimation of the coverage interval when 10 and 20 digits are used. When 10 digits are used, this underestimation is of the order of 7% for short distances (4 m) and 12% for large distances (20 m). These results show that the coverage interval value increases when distance increases (around 30%).

## Conclusions

Multilateration is a measurement method with numerous applications in metrology and precision engineering. In this paper, the effects of the number of digits on large scale and sequential multilateration were studied. The effects of numerical errors due to the floating-point precision were analyzed. For large distance measurements (20 meters) in a monitored environment (temperature, humidity, atmospheric pressure), the multilateration technique allowed to obtain a distance coverage interval around ±0.046mm. Increasing the digit number decreases the numerical error of the computed quantities and allows better estimates of the coverage intervals. To conclude, the use of multi-precision libraries is recommended to control the uncertainty propagation during multilateration computation.

## Acknowledgments

The authors gratefully acknowledge the contributions of Elias Rechreche during his internship. Measuring instruments used in the experiments were funded by the European Community, French Ministry of Research and Education, Pays d'Aix Conurbation Community, Aix Marseille University.

## References

[1] Puttock, M. J., (1978), Large-Scale Metrology, CIRP Annals, 2711(1), 351-6.

[2] Estler, W. T., Edmundson, K. L., Peggs, G. N., & Parker, D. H. (2002). Large-scale metrology–an update. CIRP annals, 51(2), 587-609.

[3] Schmitt, R. H., Peterek, M., Morse, E., Knapp, W., Galetto, M., Härtig, F., ... & Estler, W. T. (2016). Advances in large-scale metrology–review and future trends. CIRP Annals, 65(2), 643-665.

[4] Schwenke, H., Knapp, W., Haitjema, H., Weckenmann, A., Schmitt, R., & Delbressine, F. (2008). Geometric error measurement and compensation of machines—an update. CIRP Annals, 57(2), 660-675.

[5] Maropoulos, P. G., Muelaner, J. E., Summers, M. D., & Martin, O. C. (2014). A new paradigm in large-scale assembly—research priorities in measurement assisted assembly. The International Journal of Advanced Manufacturing Technology, 70(1-4), 621-633.

[6] Schwenke, H., Franke, M., Hannaford, J., & Kunzmann, H. (2005). Error mapping of CMMs and machine tools by a single TI. CIRP annals, 54(1), 475-478.

[7] Muralikrishnan, B., Phillips, S., & Sawyer, D. (2016). Laser trackers for large-scale dimensional metrology: A review. Precision Engineering, 44, 13-28.

[8] Norrdine, A. (2012, November). An algebraic solution to the multilateration problem. In Proceedings of the 15th international conference on indoor positioning and indoor navigation, Sydney, Australia (Vol. 1315).

[9] Gao, W., Kim, S. W., Bosse, H., Haitjema, H., Chen, Y. L., Lu, X. D., ... & Kunzmann, H. (2015). Measurement technologies for precision positioning. CIRP Annals, 64(2), 773-796.

[10] Chen, H., Tan, Z., Shi, Z., Song, H., & Yan, H. (2016). Optimization Method for Solution Model of Laser Tracker Multilateration Measurement. Measurement Science Review, 16(4), 205-210.

[11] Camboulives, M., Lartigue, C., Bourdet, P., & Salgado, J. (2016). Calibration of a 3D working space by multilateration. Precision Engineering, 44, 163-170.

[12] Zhang, D., Rolt, S., & Maropoulos, P. G. (2005). Modelling and optimization of novel laser multilateration schemes for high-precision applications. Measurement Science and Technology, 16(12), 2541.

[13] Wang, H., Cai, Y., Li, T., Wang, L., & Li, F. (2016). Application of genetic algorithm to multilateration measurement of the volumetric error in machine tools. Advances in Mechanical Engineering, 8(9), 1687814016666450.

[14] Santolaria, J., Majarena, A. C., Samper, D., Brau, A., & Velázquez, J. (2014). Articulated arm coordinate measuring machine calibration by laser tracker multilateration. The Scientific World Journal, 2014.

[15] Aguado, S., Santolaria, J., Samper, D., & Aguilar, J. J. (2013). Influence of measurement noise and laser arrangement on measurement uncertainty of laser tracker multilateration in machine tool volumetric verification. Precision Engineering, 37(4), 929-943.

[16] Conte, J., Majarena, A. C., Aguado, S., Acero, R., & Santolaria, J. (2016). Calibration strategies of laser trackers based on network measurements. The International Journal of Advanced Manufacturing Technology, 83(5-8), 1161-1170.

[17] Hongfang, C. H. E. N., Jiang, B., Shi, Z., Sun, Y., Song, H., & Tang, L. (2018). Uncertainty modelling of the spatial coordinate error correction system of the CMM based on laser tracer multi-station measurement. Measurement Science and Technology.

[18] Wang, H., Shao, Z., Fan, Z., & Han, Z. (2019). Configuration optimization of laser tracker stations for position measurement in error identification of heavy-duty machine tools. Measurement Science and Technology.

[19] Mutilba, U., Yagüe-Fabra, J. A., Gomez-Acedo, E., Kortaberria, G., & Olarra, A. (2018). Integrated multilateration for machine tool automatic verification. CIRP Annals, 67(1), 555-558.

[20] Ibaraki, S., Kudo, T., Yano, T., Takatsuji, T., Osawa, S., & Sato, O. (2015). Estimation of three-dimensional volumetric errors of machining centers by a TI. Precision Engineering, 39, 179-186.

[21] Xusheng Z., Lianyu Z., & Xiaojun T. (2016). Configuration optimization of laser tracker stations for large-scale components in non-uniform temperature field using Monte-Carlo method. Procedia CIRP, 56, 261-266.

[22] Takatsuji, T., Koseki, Y., Goto, M., & Kurosawa, T. (1998). Restriction on the arrangement of laser trackers in laser trilateration. Measurement Science and Technology, 9 (8), 1357.

[23] Linares, J. M., Goch, G., Forbes, A., Sprauel, J. M., Clément, A., Haertig, F., & Gao, W. (2018). Modelling and traceability for computationally-intensive precision engineering and metrology. CIRP Annals, 67(2), 815-838.

[24] Edlén, B. (1966). The refractive index of air. Metrologia, 2(2), 71.

[25] Bönsch, G., & Potulski, E. (1998). Measurement of the refractive index of air and comparison with modified Edlén's formulae. Metrologia, 35(2), 133.